\documentclass[preprint,preprintnumbers]{revtex4}

\usepackage{graphicx}

\usepackage{amssymb}

\usepackage{mathrsfs}

\begin{document}

\title{THE ANDREEV  CROSSED REFLECTION-A MAJORANA PATH INTEGRAL APPROACH  }

\author{D. Schmeltzer}

\affiliation{Physics Department, City College of the City University of New York,  
New York, New York 10031, USA}


\begin{abstract}

\noindent
We investigate  the effect of the Majorana Fermions which are formed at the boundary of   a p-wave  superconductor.  When the Majorana  overlapping energy is finite  we construct the scattering matrix $\mathbf{S}$  by   maping   the Majorana  zero mode  to     Fermions  for which coherent states are defined and a path integral  is obtained .  The path integral is used   to compute     the scattering   matrix  in terms  of the electrons in the leads .  This method is suitable for  computing  the conductivity.  We investigate a  chiral Majorana  Hamiltonian and show that in the   absence of vortices the conductivity  vanish.   We compute  the  conductivity for  p wave superconductor coupled  to two  metallic leads we show that  when the overlapping  energy  between the two  Majorana fermions is finite  the Andreev Crossed reflection conductance is  finite.


\end{abstract}



\maketitle


\noindent
\textbf{1.  INTRODUCTION}

\vspace{0.2 in}

\noindent
At the surface of a topological insulator electrons  carry a Berry  phase  of $\pi$, in the presence of an  attractive interactions superconductivity is induced.    For a low level  doping we obtain a p-wave   topological superconductor. 
 Majorana Fermions   appear on the surface of a topological insulator  in a region where the  chemical potential $\mu_{eff.}(\vec{r})$  changes sign . We consider the  effect of the Majorana modes on the  $p-wave$ superconductor \cite{Read,Ivanov,Alicea,Oreg}.
When  two metallic leads are attached to  the superconductor, the Majorana fermion  induces resonant Andreev  reflection  \cite{Ng} or crossed  Majorana    Andreev  reflection. With increasing doping, a regular superconductor  is expected with the  Andreev conductance of the order of $\frac{e^2}{h}(\frac{\Gamma}{\Delta})^2$ ($\Gamma$   is the tunneling width   and  $\Delta$ is the superconducting gap), which is much smaller than the Andreev conductance carried  by the Majorana fermions.
 The  phenomena of  Andreev reflection and  crossed  Andreev reflection    can be understood from the general properties of the   scattering  $\mathbf{S}$ \cite{Beenaker,Ng,Fisher,Beri,Buttiker,Fan,Flensberg}.
The scattering matrix $\mathbf{S}$  is computed  using the continuity equations and  the unitarity properties.
 For finite  Majorana energies,  it is difficult to obtain  the scattering matrix $\mathbf{S}$.   For such cases it is preferable     to represent the scattering matrix    $\mathbf{S}$ as a Dyson series  $\mathbf{S}=T\Big[e^{- i \int_{-\infty}^{\infty}\,dt'  H_{eff.}(t')}\Big] $ \cite{Weinberg}, expressed  in terms of the leads   Hamiltonian. This is obtained  by  integrating   the Majorana fermions.
This approach has the advantage of taking account the renormalization effect for the  tunneling matrix element.  Experimentally the tunneling  for the differential conductance is in disagreement with the quantized  values \cite{Yazdani}.
\noindent
\textbf{The purpose of this paper is to introduce    the  scattering matrix    $\mathbf{S}$ as a Dyson series  $\mathbf{S}=T\Big[e^{- i \int_{-\infty}^{\infty}\,dt'  H_{eff.}(t')}\Big] $. Using the scattering matrix  we will  compute the differential conductance for  different  cases considered in the literature}
\noindent
The plan of the paper is as following:
In sec. 2. we formulate the problem in terms of the leads and the fermionic representation of the Majorana fermions.
 In sec. 3. we consider a superconducting island deposited on the surface of a three dimensional topological  insulator. The area outside the superconductor is gaped  by a ferromagnetic material. We demonstrate that in   the absence of vortices  the conductivity between the metallic leads vanish. In sec.4. we consider two Majorana fermions coupled to two leads and compute the Andreev crossed reflection for the p-wave superconductor .  Sec .5. is devoted to conclusions.

\vspace{0.2 in}
\noindent
\textbf{2.  MAJORANA  FERMIONS FOR  A P-WAVE  SUPERCONDUCTOR}

\vspace{0.2 in}
\noindent
We consider a p-wave superconductor  described by the $\Psi_{\sigma}(x,y)$   the Bogoliubov -de Genes  fermion operator. At the boundary between the superconductor and the metallic leads  Majorana zero modes are formed (the chemical potential changes sign).  The Bogoliubov -de Genes operator contains also  the zero modes  given  by the operator   $\hat{C}_{0}(\vec{r}) $.
The coupling of the p-wave  superconductor  to the  two leads is given by,
\begin{equation}
 H_{t}=t\sum_{\sigma=\uparrow,\downarrow}\int\,dy\Big[d^{\dagger}_{\sigma}(x=-\frac{L}{2},y)\Psi_{\sigma}(x=-\frac{L}{2},y)+d^{\dagger}_{\sigma}(x=\frac{L}{2},y)\Psi_{\sigma}((x=\frac{L}{2},y)+h.c.\Big]
\label{equation}
\end{equation}
$d^{\dagger}_{\sigma}(x=-\frac{L}{2},y)$,$d_{\sigma}(x=-\frac{L}{2},y)$ are the fermions in the left lead and $d^{\dagger}_{\sigma}(x=\frac{L}{2},y)$,$d_{\sigma}(x=\frac{L}{2},y)$ represent the fermions in the right lead.

\noindent
 I the presence of  the  Majorana fermions    we replace   the Bogoliubov -de Genes operator    $\Psi_{\sigma}(x,y)$ by the zero mode part $\hat{C}_{0}(\vec{r})$.

\noindent
 For an \textbf{even} number of Majorana fermions, we replace  the zero mode   $\hat{C}_{0}(\vec{r})$ by  the representation : 
\begin{equation}
\hat{C}_{0}(\vec{r})=\sqrt{2}\sum_{a=1}^{n}\Big[\hat{\gamma}_{2a-1}F_{2a-1}(\vec{r})+\hat{\gamma}_{2a}F_{2a}(\vec{r})\Big]
\label{spinor}
\end{equation}
 The spinors are given by   $F_{2a}(\vec{r})$ and  $F_{2a-1}(\vec{r})$
\begin{eqnarray}
&&F_{2a}(\vec{r})=
\Big[\frac{1}{\sqrt{i}}e^{\frac{i }{2}\phi_{2a}},\frac{1}{\sqrt{-i}}e^{\frac{-i }{2}\phi_{2a}}\Big]^{T}\frac{f_{2a}(r)}{\sqrt{r}}
\nonumber\\&&
F_{2a-1}(\vec{r})=
\Big[\frac{1}{\sqrt{i}}e^{\frac{i }{2}\phi_{2a-1}},\frac{1}{\sqrt{-i}}e^{\frac{-i }{2}\phi_{2a-1}}\Big]^{T}\frac{f_{2a-1}(r)}{\sqrt{r}}\nonumber\\&&
\end{eqnarray} 
The    two component spinors  are  localized at the positions $\vec{r}=\vec{R}_{2a-1}$ and $\vec{r}=\vec{R}_{2a}$.
We introduce the fermion operators $ \zeta^{\dagger}_{a}$ and $\zeta_{a}$ , $a=1,2,3...n$. The transformation between the two representation is given by: 
  $\hat{\gamma}_{2a-1}=\frac{1}{\sqrt{2}}\Big[\zeta^{\dagger}_{a}+\zeta_{a}\Big]$, 
$\hat{\gamma}_{2a}=\frac{1}{i\sqrt{2}}\Big[\zeta^{\dagger}_{a}-\zeta_{a}\Big]$, $a=1,2,3...n$

\noindent 
The overlap between different Majorana fermions will  introduce  the   energy $\epsilon_{a}$   for the   Majorana Hamiltonian: $H^{(Majorana)}=\sum_{a=1}^{a=n}i\epsilon_{a}\hat{\gamma}_{2a-1}\hat{\gamma}_{2a}=
\sum_{a=1}^{a=n}\epsilon_{a}\zeta^{\dagger}_{a}\zeta_{a}$.
This allows to obtain     the  Majorana  action,
\begin{equation}
\mathbf{S}^{(Majorana)}= \sum_{a=1}^{a=n}\int\,dt  \Big[\zeta^{\dagger}_{a}(i\partial_{t})\zeta_{a}-\epsilon_{a}\zeta^{\dagger}_{a}\zeta_{a}\Big]
\label{action}
\end{equation}
The action in Eq.$(4)$ allows for the construction of the  integral for the Majorana fermions  which will be used for computing the conductivity.
\noindent For an \textbf{odd} number of Majorama Fermions  we will have for the  $2 n+1$  Majorana  an unpaired Fermionic  ,we can choose for  $\hat{\gamma}_{n+1}=\frac{1}{\sqrt{2}}\Big[\zeta^{\dagger}_{n+1}+\zeta_{n+1}\Big]$\textbf{ or}   $\hat{\gamma}_{n+1}=\frac{1}{i\sqrt{2}}\Big[\zeta^{\dagger}_{n+1}-\zeta_{n+1}\Big]$.

\vspace{0.2 in}
\noindent
\textbf{3. A CHIRAL MAJORANA  FERMION COUPLED  TO  TWO LEADS  }  

\vspace{0.2 in}
\noindent
We consider a grounded   superconducting island of radius $R$ deposited on the surface of a three dimensional topological  insulator. The area outside the superconductor is gaped  by a ferromagnetic material. 
We will attach the superconducting island to two leads at $\theta=0$ (left lead)  and $\theta=\pi$ (right lead). We will show that in the absence of vortices the left lead is effectively  not coupled to the right lead  and  therefore the conductance vanish.

\vspace{0.1 in}
\noindent
\textbf{3.1. No vortex in  the superconductor}

\vspace{0.1 in}
\noindent
 The Hamiltonian at  the interface  is described by a chiral Majorana Hamiltonian.
\begin{equation}
H^{Majorana}=\frac{\hbar v}{R}\displaystyle\oint_{0}^{2\pi}\,d\theta
\hat{\gamma}(\theta)(-i\partial_{\theta})\hat{\gamma}(\theta) 
\label{Maj}
\end{equation}
We replace the Majorana fermion  $\hat{\gamma}(\theta)$  by regular fermions $C(\theta) $ and $C^{\dagger}(\theta)$ , $\hat{\gamma}(\theta)=\frac{1}{\sqrt{2}}[C^{\dagger}(\theta)+C(\theta)]$ and expand the fermion in angular momentum states: $C(\theta)=\sum_{l}e^{il\theta}C_{l}$. The Majorana Hamiltonian takes the Bogoliubov-de Genes form: 
\begin{equation}
H^{Majorana}=\frac{h v}{R}\sum_{l>0}\Big[(C^{\dagger}_{l}C_{l}-C^{\dagger}_{-l}C_{-l})l+(C_{l}C_{-l}-C^{\dagger}_{l}C^{\dagger}_{-l})l\Big]
\label{ew}
\end{equation}
The  Bogoliubov-de Genes eigenvalues  for the Hamiltonian  in   Eq.$(6)$ are: $\epsilon_{\lambda=0}=0$ and $\epsilon_{\lambda \neq 0}=\frac{h v}{R}2l$.  The eigenspinors  are ,$\chi(l)$( for the zero eigenvalue ) and $\eta(l)$ ( for the  non zero eigenvalues).
\begin{eqnarray}
&&\chi(l)=\frac{1}{\sqrt{2}}\Big[C^{\dagger}(l)+C(-l)]\nonumber\\&&
\eta(l)=\frac{1}{\sqrt{2}}\Big[C(l)-C^{\dagger}(-l)\Big], 
\vspace{0.1 in} \eta^{\dagger}(l)=\frac{1}{\sqrt{2}}\Big[C^{\dagger}(l)-C(-l)\Big]\nonumber\\&&
H^{Majorana}=\frac{h v}{R}\sum_{l}(2l)\eta^{\dagger}(l)\eta(l)\nonumber\\&&
\end{eqnarray} 
The tunneling Hamiltonian is given by
\begin{eqnarray}
&&H_{t}=g\Big[d^{\dagger}_{1}(\theta=0)e^{i\frac{\phi_{1}}{2}}-d_{1}(\theta=0)e^{-i\frac{\phi_{1}}{2}})\Big]\hat{\gamma}(\theta=0)+
g\Big[ d^{\dagger}_{2}(\theta=\pi)e^{i\frac{\phi_{1}}{2}}-d_{2}(\theta=\pi)e^{-i\frac{\phi_{1}}{2}})\Big]\hat{\gamma}(\theta=\pi),\nonumber\\&&
\end{eqnarray}
We substitute the eigenspinor ,$\chi(l)$ and  $\eta(l)$ and find:
\vspace{0.1 in}
\begin{eqnarray}
&&H_{t}=g(d^{\dagger}_{1}(\theta=0)e^{i\frac{\phi_{1}}{2}}-d_{1}(\theta=0)e^{-i\frac{\phi_{1}}{2}})\sum_{l}\chi(l)+\nonumber\\&&
g( d^{\dagger}_{2}(\theta=\pi)e^{i\frac{\phi_{1}}{2}}-d_{2}(\theta=\pi)e^{-i\frac{\phi_{1}}{2}})(\sum_{l(even)}\chi(l)-\sum_{l(odd)}\eta(l))\nonumber\\&&
\end{eqnarray}
The  Hamiltonian $H_{t}$ in Eq.$(9)$ is independent from $\chi(l)$ .  The integration of of the Majorana fermions   in Eqs.$(7-10)$ will give a scattering matrix.We  find that the scattering matrix  depends only on the\textbf{ right lead}! The left lead  $\Big[d^{\dagger}_{1}(\theta=0)e^{i\frac{\phi_{1}}{2}}-d_{1}(\theta=0)e^{-i\frac{\phi_{1}}{2}}\Big]$  which couples to $\chi(l)$  will not appear in the scattering matrix!
As a result  the  cross- Andreev conductance  will vanish.

\vspace{0.1 in}
\noindent
\textbf{3.2. A  vortex  inside the superconductor}

\vspace{0.1 in}
\noindent
When a vortex is added to the case considered in case given in $3.1$
we need to add the impurity Hamiltonian:
\begin{eqnarray}
&&H_{imp.}=\displaystyle\oint_{0}^{2\pi}\,d\theta
t(\theta;\vec{r}_{0})\hat{\gamma}(\theta)Y_{vortex}(\vec{r}_{0})\nonumber\\&&
\end{eqnarray}
$Y_{vortex}(\vec{r})$ is the Majorana  vortex which  couple with the strength $t(\theta;\vec{r}_{0})$ to the chiral Majorana fermion $\hat{\gamma}(\theta;\vec{r}_{0})$. Due to this coupling  $t(\theta;\vec{r}_{0})$ the  two leads will be coupled and  the cross- Andreev conductance  will be finite .
The exact result of the Andreev   conductance will depend on the details of the coupling $\hat{\gamma}(\theta;\vec{r}_{0})$.

\vspace {0.2 in}
\noindent
\textbf{4. A PAIR OF TWO MAJORANA FERMIONS COUPLED TO TWO LEADS}  

\vspace{0.2 in}
\noindent
We consider a grounded  p-wave topological superconductor  attached to two leads.  Close to the leads due to the boundary  condition  the p-wave superconductor has to Majorana modes.
\noindent
 We will  compute the \textbf{Crossed Andreev  Reflection}  a process where  \textbf{an incoming electron from lead $1$ is turned into an outgoing hole in lead $2$. In this case a  single electron at each lead   is tunneling into  superconductor to form a Cooper pair}.  
 We consider two   half vortices localized in the superconductor  at  $\vec{r}=\vec{R}_{1}=[x_{1}\approx \frac{-L}{2},y=0]$  and  $\vec{r}=\vec{R}_{2}=[x_{2}\approx \frac{L}{2},y=0]$. For this case, we have  for the zero modes,
\begin{equation}
\hat{C}_{0}(\vec{r})=\hat{C}^{\dagger}_{0}(\vec{r})=\sqrt{2}\Big[\hat{\gamma}_{1}F_{1}(\vec{r})+\hat{\gamma}_{2}F_{2}(\vec{r})\Big]
\label{newleads}
\end{equation}
where $\hat{\gamma}_{1}$ and $\hat{\gamma}_{2} $ are the two Majorana operators.
\noindent We attach the two leads at  $[x=-\frac{L}{2},y=0]$ and 
$[x=\frac{L}{2},y=0]$,  and  due to the non-locality of the  spinors $ F_{1}(\vec{r})$  , $F_{2}(\vec{r})$ the Majorana fermions couples to the fermions in the two leads .   We consider a situation where the  two  Majorana  fermions      overlap  with  energy $\epsilon_{0}$.  Using the energy  $\epsilon_{0}$ we construct the  $ H^{(Majorana)}$ Hamiltonian.       The tunneling Hamiltonian  between the leads  and the Majorana fermions  is given by the Hamiltonian  $ H_{t}$: 
\begin{eqnarray}
&&H_{t}=g\Big[d^{\dagger}_{1}(x=-\frac{L}{2},0)e^{i\frac{\phi_{1}}{2}}-d_{1}(x=-\frac{L}{2},0)e^{-i\frac{\phi_{1}}{2}}\Big]\hat{\gamma}_{1}+ \Big[d^{\dagger}_{2}(x=\frac{L}{2},0)e^{i\frac{\phi_{2}}{2}}-d_{2}(x=\frac{L}{2},0)e^{-i\frac{\phi_{2}}{2}}\Big]\hat{\gamma}_{2}\nonumber\\&&
H^{(Majorana)}=i\epsilon_{0}\hat{\gamma}_{1}\hat{\gamma}_{2}\nonumber\\&&
\end{eqnarray}
$d^{\dagger}_{1}(x=-\frac{L}{2},0)$, $ d_{1}(x=-\frac{L}{2},0)$ are the fermions  in the left lead and  $d^{\dagger}_{2}(x=-\frac{L}{2},0)$, $ d_{2}(x=-\frac{L}{2},0)$ are the fermions in the right lead.
 $\epsilon_{0}$ describes the  overlapping between  the two Majorana Fermions (two independent half vortices). The two   vortices  are  localized at  positions $\vec{R}_{1}$ , $\vec{R}_{2}$ and   their wave functions  are  non-orthogonal.
We replace the two Majorana  Fermions    with a single  fermion ,  $\hat{\gamma}_{1}=\frac{1}{\sqrt{2}}\Big[\zeta^{\dagger}+\zeta\Big]$  and $\hat{\gamma}_{2} =\frac{1}{i\sqrt{2}}\Big[\zeta^{\dagger}-\zeta\Big]$. The tunneling     Hamiltonian is given in terms of leads operators  $ V^{\dagger}$ and $V$ form:
\begin{equation}
H_{t}=\frac{g}{\sqrt{2}}\Big[V^{\dagger} \zeta +\zeta^{\dagger} V \Big]
\label{tunneling}
\end{equation}
The operators  $ V^{\dagger}$ and $V$   are expressed in terms of the one dimensional leads:
\begin{eqnarray}
&&V^{\dagger}=  \frac{1}{i\sqrt{2}}\Big[e^{i\frac{\phi_{1}}{2}}d^{\dagger}_{1}(x=-\frac{L}{2},0)-e^{-i\frac{\phi_{1}}{2}}d_{1}(x=-\frac{L}{2},0)\Big]+ \frac{1}{\sqrt{2}}\Big[e^{i\frac{\phi_{2}}{2}}d^{\dagger}_{2}(x=\frac{L}{2},0)-e^{-i\frac{\phi_{2}}{2}}d_{2}(x=\frac{L}{2},0)\Big]\nonumber\\&&
V= - \frac{1}{i\sqrt{2}}\Big[e^{i\frac{\phi_{1}}{2}}d^{\dagger}_{1}(x=-\frac{L}{2},0)-e^{-i\frac{\phi_{1}}{2}}d_{1}(x=-\frac{L}{2},0)\Big]
 + \frac{1}{\sqrt{2}}\Big[e^{i\frac{\phi_{2}}{2}}d^{\dagger}_{2}(x=\frac{L}{2},0)-e^{-i\frac{\phi_{2}}{2}}d_{2}(x=\frac{L}{2},0)\Big]\nonumber\\&&
\end{eqnarray}
The action for this case is given by:
\begin{equation}
S=\int\,dt\Big[\frac{1}{2}\Big(\zeta^{\dagger}(t)(i\partial_{t})\zeta(t)+\zeta(t)(i\partial_{t})\zeta^{\dagger}(t)\Big) -\epsilon_{0}\zeta^{\dagger}\zeta-H_{t}(t)\Big]
\label{action}
\end{equation}
 Using  the Grassman integration \cite{Nakahara} (see Eq.$(1.191)$ in Nakahara) for the   Majorana  Fermions   $\zeta^{\dagger}$,$\zeta$  we obtain   the effective Hamiltonian $ H_{eff.}(t)$  for  the leads: 
\begin{eqnarray} 
&&H_{eff.}(t)=(-ig^2) \int_{0}^{\infty}\,d \tau\Big[V^{\dagger}(t)e^{i\epsilon_{0} \tau}V(t-\tau)\Big];\nonumber\\&&
V^{\dagger}(t) V(t-\tau)\equiv\Big(d^{\dagger}_{1}(t)e^{-i\frac{\phi_{1}}{2}}-d_{1}(t)e^{i\frac{\phi_{1}}{2}}+ id^{\dagger}_{2}(t)e^{-i\frac{\phi_{2}}{2}}-id_{2}(t)e^{i\frac{\phi_{2}}{2}}\Big)\cdot
\nonumber\\&&\Big(-d^{\dagger}_{1}(t-\tau)e^{-i\frac{\phi_{1}}{2}}+d_{1}(t-\tau)e^{i\frac{\phi_{1}}{2}}+ id^{\dagger}_{2}(t-\tau)e^{-i\frac{\phi_{2}}{2}}-id_{2}(t-\tau)e^{i\frac{\phi_{2}}{2}}\Big)\nonumber\\&&
\end{eqnarray}
For the electrons  in the leads, we use  the  right ($R$) and left ($L$)  movers representation. $d_{1}(x=-\frac{L}{2},0)\equiv d_{1}(t)$  and $d^{\dagger}_{1}(x=-\frac{L}{2},0)\equiv d^{\dagger}_{1}(t)$  are the electrons  in the left lead $(1)$ and  $d_{2}(x=\frac{L}{2},0)\equiv d_{2}(t)$  and $d^{\dagger}_{2}(x=\frac{L}{2},0)\equiv  d^{\dagger}_{2}(t)$ are the eletrons in the right lead  $(1)$.
\begin{equation} 
d_{1}(t)=R_{1}(t)e^{-ik_{F}\frac{L}{2}}+
L_{1}(t)e^{ik_{F}\frac{L}{2}}; 
d_{2}(t)=R_{2}(t)e^{ik_{F}\frac{L}{2}}+
L_{2}(t)e^{-ik_{F}\frac{L}{2}} 
\label{eqw}
\end{equation}
We apply on the left lead a voltage  $V/2$, and on the  right lead a voltage $-V/2$.
As a result, we obtain for each lead, two Green's functions.
For the left lead (1) we have  $\mathbf{G}_{0}^{1,R}(E,\omega)$ (right mover) and $\mathbf{G}_{0}^{1,L}(E,\omega)$ (left mover).
\begin{eqnarray}
&&\mathbf{G}_{0}^{1,R}(E,\omega)=\frac{\Theta(E-\frac{eV}{2})}{\omega-(E-\frac{eV}{2})+i0}+\frac{\Theta(-E+\frac{eV}{2})}{\omega-(E-\frac{eV}{2})-i0}\nonumber\\&&
\mathbf{G}_{0}^{1,L}(E,\omega)=\frac{\Theta(-E+\frac{eV}{2})}{\omega+(E-\frac{eV}{2})+i0}+\frac{\Theta(E-\frac{eV}{2})}{\omega+(E-\frac{eV}{2})-i0)}\nonumber\\&&
\end{eqnarray}
Similarly, for the right (2) lead we have
\begin{eqnarray}
&&\mathbf{G}_{0}^{2,R}(E,\omega)=\frac{\Theta(E+\frac{eV}{2})}{\omega-(E+\frac{eV}{2})+i0}+\frac{\Theta(-E-\frac{eV}{2})}{\omega-(E+\frac{eV}{2})-i0}\nonumber\\&&
\mathbf{G}_{0}^{2,L}(E,\omega)=\frac{\Theta(-E-\frac{eV}{2})}{\omega+(E+\frac{eV}{2})+i0}+\frac{\Theta(E+\frac{eV}{2})}{\omega+(E+\frac{eV}{2})-i0}\nonumber\\&&
\end{eqnarray}
\noindent
$\Theta(x)$ is the step function which is zero  for $ x<0$ and one for $x\geq 0$).
The current in the  leads is given by:
$J(x=-\frac{L}{2};\frac{V}{2})=ev(N_{0}^{1,R}-N_{0}^{1,L})=J(x=\frac{L}{2};\frac{-V}{2})=ev(N_{0}^{2,R}-N_{0}^{2,L})$. $v$ is then  electron velocity in both  leads,   $N_{0}^{1,R} -N_{0}^{1,L}$ is the current density in the left $(1)$ lead,  and $N_{0}^{2,R}-N_{0}^{2,L}$ is the current density in the right lead $(2)$ \cite{Schulz}.
In order to compute the current, we will compute the Green's functions . The Green's function will be computed perturbatively    using  the effective coupling  to the leads $ H_{eff.}(t)$ given in Eq.$(16)$.   $ H_{eff.}(t)$ is represented in terms of one dimensional fermions given in  Eq.$(17)$ we have for each  lead  $\mathbf{right}$ ($R_{1}$, $R_{2}$) and $\mathbf{left}$ ($L_{1}$, $L_{2}$)  fermions.  The perturbation theory is controlled  by the coupling constant $g^2$ . 
\textbf{We will compute perturbatively  the Green's function   $\mathbf{G}^{1,R}(E,\omega;\frac{eV}{2})$, $\mathbf{G}^{1,L}(E,\omega;\frac{eV}{2})$ (left leads) and 
$\mathbf{G}^{2,R}(E,\omega;-\frac{eV}{2})$, $\mathbf{G}^{2,L}(E,\omega;-\frac{eV}{2})$ (right leads) }(the index $1$ and $2$ represent  the leads and $L$  and $R$ represents the left and right fermions.This Green's function contains the contributions of  the \textbf{ particles-holes, particles-particles, and  holes-holes in the same and opposite leads}.
From the Green's function we extract  the 
  self energies for each lead  and each mover,
$\mathbf{\Sigma^{1,R}}(\omega)$, $\mathbf{\Sigma^{1,L}}(\omega)$ and  $\mathbf{\Sigma^{2,R}}(\omega)$, $\mathbf{\Sigma^{2,L}}(\omega)$.
We find, to order  $g^4$,  the self energies:
\begin{eqnarray}
&&\mathbf{\Sigma^{1,R}}(\omega)=-2T(\omega,\omega_{0})\frac{g^4}{v}Ln\Big(\frac{1+\frac{\omega-\frac{eV}{2}}{\Lambda}}{1-\frac{\omega-\frac{eV}{2}}{\Lambda}}\Big)+i2T(\omega,\omega_{0})\frac{g^4}{v}\mathbf{sgn}(\omega)\nonumber\\&&
\mathbf{\Sigma^{1,L}}(\omega)=-2T(\omega,\omega_{0})\frac{g^4}{v}Ln\Big(\frac{1+\frac{\omega-\frac{eV}{2}}{\Lambda}}{1-\frac{\omega-\frac{eV}{2}}{\Lambda}}\Big)-i2T(\omega,\omega_{0})\frac{g^4}{v}\mathbf{sgn}(\omega)\nonumber\\&&
T(\omega,\omega_{0})=\frac{1}{\Gamma^2_{0}+(\omega+\omega_{0})^2}\nonumber\\&&
\end{eqnarray}
Where $\Lambda$ is the band with, $\Gamma_{0}$ is a damping factor which is induced at high momenta, and   $\hbar\omega_{0}=\epsilon_{0}$ is the Majorana energy. The \textbf{imaginary} part of the self energy obeys \textbf{ $Im. \mathbf{\Sigma^{1,L}}(\omega)=-Im. \mathbf{\Sigma^{1,R}}(\omega)$} and the \textbf{real part of the self energy} obeys \textbf{ $\Re \mathbf{\Sigma^{1,L}}(\omega)=\Re \mathbf{\Sigma^{1,R}}(\omega)\equiv\mathbf{\Sigma^{1}}(\omega)$}.
The Green's funtions are  given in terms of the  self energies:
\begin{eqnarray}
&&\mathbf{G}^{1,R}(E,\omega;\frac{eV}{2})=\Big(\omega-(E-\frac{eV}{2})-\mathbf{\Sigma^{1,R}}(\omega)\Big)^{-1}\nonumber\\&&
\mathbf{G}^{1,L}(E,\omega;\frac{eV}{2})=\Big(\omega+(E-\frac{eV}{2})-\mathbf{\Sigma^{1,L}}(\omega)\Big)^{-1}\nonumber\\&&
\end{eqnarray}
The real part of the self energy is used to compute the wave function renormalization function  $\mathbf{Z}$.
\begin{eqnarray}
&&\Big(1-\partial_{\omega}\mathbf{\Sigma^{1}}(\omega)\Big)|_{\omega=0}=\mathbf{Z}^{-1}\nonumber\\&&
\mathbf{Z}^{-1}=\Big[1+\frac{\hat{\Gamma}}{\Lambda}\Big(\frac{1}{1-\frac{eV}{2\Lambda}}\Big]\nonumber\\&&
\hat{\Gamma}=\frac{4T(\omega=0,\omega_{0})g^4}{v }= \frac{4g^4}{v (\Gamma^2_{0}+\omega_{0}^2)}\nonumber\\&&
\end{eqnarray}
The tunneling current at the left  leads will be given by  $I(V)=ev(N^{1,R}-N^{1,L})$  (which replaces  $I(V=0)=ev(N_{0}^{1,R}-N_{0}^{1,L})$ the expression for zero voltage)   in terms of the  renormalized Green's function.
\begin{eqnarray}
&&I(V)=ev\Big(N^{1,R}-N^{1,L}\Big)\nonumber\\&&
=\frac{e}{h}(-i)\int_{-\Lambda}^{\Lambda}\,dE\int_{-\infty}^{\infty}\,\frac{d\omega} {2\pi}e^{i\omega 0^{+}}\Big[\Big(\frac{\mathbf{Z}\Theta(\omega)}{\omega-E \mathbf{Z} -\frac{eV}{2}\mathbf{Z}-i \hat{\Gamma}\mathbf{Z}}+\frac{\mathbf{Z}\Theta(-\omega)}{\omega-E \mathbf{Z} +\frac{eV}{2}\mathbf{Z}+i \hat{\Gamma}\mathbf{Z}}\Big)\nonumber\\&&
-\Big(\frac{\mathbf{Z}\Theta(\omega)}{\omega-E \mathbf{Z} -\frac{eV}{2}\mathbf{Z}+i \hat{\Gamma}\mathbf{Z}}+\frac{\mathbf{Z}\Theta(-\omega)}{\omega-E \mathbf{Z} +\frac{eV}{2}\mathbf{Z}-i \hat{\Gamma}\mathbf{Z}}\Big)\Big]\nonumber\\&&
\end{eqnarray}
\noindent  We will use $I(V)=ev\Big(N^{1,R}-N^{1,L}\Big)$ to   evaluate the \textbf{differential conductance for the Crossed Andreev reflection}
$\frac{d I(V)}{dV}$.  Due to the nonlinearity of the effective action, we will use   the scaling equations  \cite{Weinberg, Shankar, Schulz,Boyanovsky,davidimpurity}  for the coupling constant $g^2$. The scaling of    $g^2$ determines  the width $\hat{\Gamma}$. We find  the Renormalization Group  equation for the width $\hat{\Gamma}$,
$\frac{d\hat{\Gamma}}{dl}=-const.\hat{\Gamma}^{2}$ with $ l=log\Big[ \frac{1}{\frac{eV}{\Lambda}}\Big]$. The solution $\hat{\Gamma}(V)$ as a function  of  $\hat{\Gamma}(V=0)$ is given by:
\begin{equation}
\hat{\Gamma}(V)=\frac{\hat{\Gamma}(V=0)}{(1+const.Log[\frac{2 \Lambda}{eV}])^2}
\label{width}
\end{equation}
This solution will be used in Eq.$(23)$, where  $\hat{\Gamma}$ is replaced by  $\hat{\Gamma}(V)$.
Substituting  $\hat{\Gamma}(V)$  gives  us the result for the differential conductance $\frac{d I(V)}{d V}$,
\begin{eqnarray}
&&\frac{d I(V)}{d V}=\frac{e^2}{h}\int_{-\Lambda \mathbf{Z}}^{\Lambda \mathbf{Z}}\,d\epsilon \int_{-\infty}^{\infty}\,\frac{d\Omega} {2\pi}\Big [\frac{\hat{\Gamma}\mathbf{Z}}{(\Omega-\epsilon)^2+(\hat{\Gamma}\mathbf{Z})^2}\Big]\frac{d}{d\Omega}\Big(f_{F.D.}(\Omega+\frac{eV}{2}\mathbf{Z})+f_{F.D.}(\Omega-\frac{eV}{2}\mathbf{Z})\Big)\nonumber\\&&
\approx  \frac{e^2}{h}\int_{-\Lambda }^{\Lambda} \,d\epsilon\frac{1}{2\pi}\Big[\frac{\hat{\Gamma}(V)}{(\frac{eV}{2}-\epsilon)^2+(\hat{\Gamma}(V))^2}+\frac{\hat{\Gamma}(V)}{(\frac{eV}{2}+\epsilon)^2+(\hat{\Gamma}(V))^2}\Big] \nonumber\\&&
=\frac{e^2}{h}\frac{1}{\pi}\Big[ArcTan[\frac{\Lambda}{\Gamma(V)}(1+\frac{eV}{2\Lambda})]+ArcTan[\frac{\Lambda}{\Gamma(V)}(1-\frac{eV}{2 \Lambda})]\Big]\nonumber\\&&
\end{eqnarray}
\begin{figure}
\begin{center}
\includegraphics[width=4.0 in ]{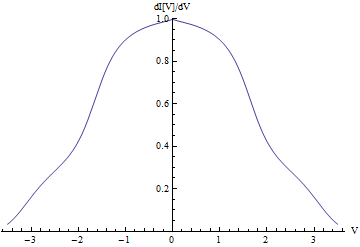} 
\end{center}
\caption{$\frac{dI}{dV} $  the differential conductivity for the Andreevv crossed reflection}
\end{figure}
\noindent 
We find that  for a pair of vortices the Andreev crossed  reflectioion obeys  $\frac{dI(V)}{dV}|_{V\rightarrow 0}  \longrightarrow \frac{e^2}{h}$ . Figure $(1)$ shows  shows the  differential $\frac{dI}{dV}$ vconductivity for the Andreev crossed reflection as a function of the voltage difference between the two leads. We observe that in the limit $V\rightarrow 0$ the maximum value for the conductance is obtained. This result follows from the scaling equation for the width given in Eq.$(24)$.

\noindent
Comparing the diferential conductivity with the experiments \cite{Yazdani} one observes that the perfect quantization is not achieved  this suggest the possibility that the width is controlled by additional operators  causing $\hat{\Gamma}(V)$ not to flow to zero when $V\rightarrow 0$.


\vspace{0.2 in}
\noindent
\textbf{5. CONCLUSIONS}

\vspace{0.2 in}

\noindent
In this paper we have introduced a new method for computing the conductance in the presence of the Majorana Fermions.
We mapp the problem of Majorana Fermions to regular Fermions for which a path integral and the Berry phase are obtained. This allows us to integrate out the Majorana Fermions and allows us to obtain     the  scattering matrix    $\mathbf{S}$ as a Dyson series  $\mathbf{S}=T\Big[e^{- i \int_{-\infty}^{\infty}\,dt'  H_{eff.}(t')}\Big] $ . Using this method we have   compute the differential conductance for  different  cases ,Achiral Majorana Fermion coupled to leads with and without an  additional  vortex  and studied the Andreev crossed reflection for a pair of Majorana coupled to two leads.
We have computed  the  differential $\frac{dI}{dV}$ vconductivity for the Andreev crossed reflection as a function of the voltage difference between the two leads. We observe that in the limit $V\rightarrow 0$  the conductance reaches the maximum value.

\end{document}